\documentclass[amstex,showpacs,preprint]{aipproc}
\usepackage{graphicx}
\usepackage{amsmath}
\usepackage{bm}
\usepackage{amsfonts}
\usepackage{amssymb}
\layoutstyle{8x11single}
\begin{document}
\title{Is Bell's theorem relevant to quantum mechanics? \\ On locality and non-commuting observables}
\author{A.\ Matzkin}
{address={Laboratoire de Spectrom\'{e}trie Physique (CNRS Unit\'{e} 5588),
Universit\'{e} Joseph-Fourier Grenoble-1, BP 87, 38402 Saint-Martin d'H\`{e}res, France}}

\begin{abstract}
Bell's theorem is a statement by which averages obtained from specific types
of statistical distributions must conform to a family of inequalities. These
models, in accordance with the EPR argument, provide for the simultaneous
existence of quantum mechanically incompatible quantities. We first recall
several contradictions arising between the assumption of a joint
distribution for incompatible observables and the probability structure of
quantum-mechanics, and conclude that Bell's theorem is not expected to be relevant
to quantum phenomena described by non-commuting observables, irrespective of
the issue of locality. Then, we try to disentangle the locality issue from
the existence of joint distributions by introducing two models accounting
for the EPR correlations but denying the existence of joint distributions.
We will see that these models do not need to resort explicitly to non-locality: the first model relies on conservation laws for ensembles, and the second model on an equivalence class by which different configurations lead to the same physical predictions.
\end{abstract}
\maketitle

\section{Introduction}

The Bell inequalities result from Bell's theorem \cite{bell,CS}.\ This
theorem is a mathematical statement, unrelated to any specific physical
theory \cite{rastall81}. Briefly put, Bell's theorem in its simplest form
tells us that average values obtained from a specific type of statistical
distribution of a variable must conform to a family of inequalities. The
specificity in question, coined under the questionable but widely used
terminology \textquotedblright local hidden variables\textquotedblright\
(LHV), is to be found in the assumptions made in the derivation of the
theorem. Its connection with quantum mechanics springs up from the dilemma
put forward by Einstein, Podolsky and Rosen \cite{EPR}: either (i) quantum
mechanics is complete or (ii) physical quantities associated with
non-commuting observables have simultaneous reality provided that locality
holds. Indeed, LHV models adopt branch (ii) of the dilemma whose main
assumption, is the 'simultaneous reality' of incompatible quantities --
locality plays the role of an auxiliary assumption to avoid action at a
distance.\ In Bell's theorem, this hazy terminology takes the form of a
precise statement: the existence of a joint probability distribution for
outcomes corresponding to incompatible observables \cite{rastall81,fine,muynck,accardi}
-- which in quantum
mechanics only exists for commuting operators.
In the first part of this note, we recall several well-known
contradictions between alternative (ii) of the EPR dilemma and quantum
mechanics.\ All these contradictions are grounded on the fact that
incompatible physical quantities require a different probabilistic structure
than the one offered by LHV models; in this sense, LHV models and the
resulting Bell inequality are not relevant to quantum mechanics. The
interesting question then is whether models complying with the
quantum-mechanical requirement of non-commutativity are, in the context of
EPR correlations, necessarily non-local. In the second part of this note,
we will introduce two types of models which will turn out to be
not necessarily non-local: in the first model, the EPR correlations can be
attributed to a conservation law (in a holistic context however). The second
model expels the locality issue from the EPR paradox by viewing a quantum
state as an equivalence class of different but equivalent field/particle
configurations.

\section{Bell-type models and incompatible physical quantities}

\subsection{Bell's theorem}

Bell's theorem (BT) is a mathematical statement giving a constraint on
certain type of probability distributions, unrelated to any specific
physical theory.\ We will nevertheless introduce the setting and the
notation in line with the two spin-1/2 particles in the singlet state
system, which is the paradigmatic application of the Bell inequalities in
quantum mechanics. We thus have two particles (formed by the fragmentation
of an initial compound system) flying apart in opposite directions.\ A
measurement, the spin projection along a chosen axis, is made on each of the
particles. Let $i=1,2$ denote the particle, $a,b$ the axis of the
measurement (the \textquotedblright parameter\textquotedblright\ of the
measurement) making respective angles $\theta_{a},\theta_{b}$ with an
arbitrarily chosen $z$ axis, and $A_{i},B_{i}$ the outcome obtained by
measuring particle $i$ along the axis $a,b,...$ Let us assume that particle $%
1$ is measured along $a$ and particle $2$ along $b$. Each measurement can
yield as possible outcomes $(A_{1},B_{2})=(\pm\frac{1}{2},\pm\frac{1}{2})$
with observed frequencies $F(A_{1},B_{2})$. The resulting expectation value
is%
\begin{equation}
E(a,b)=\sum_{A_{1},B_{2}}A_{1}B_{2}F(A_{1},B_{2})
\end{equation}
where $A_{1},B_{2}=\pm\frac{1}{2}$.

Bell's theorem arises by supposing that each measurement is actually
determined by an unknown variable $\lambda$ that completely specifies the
state of the system. $\Lambda$ denotes the set containing all the $\lambda$%
's, and $\rho(\lambda)$ the normalized distribution of the variable
corresponding to a certain state of preparation of the system. Each $\lambda$
gives rise to an outcome $(A_{1}(\lambda),B_{2}(\lambda))\ $with a
probability $p(A_{1},B_{2},\lambda)$. The observed frequencies are obtained
by averaging over $\rho(\lambda)$%
\begin{equation}
F_{\rho}(A_{1},B_{2})=\int p(A_{1},B_{2},\lambda)\rho(\lambda)d\lambda
\label{17}
\end{equation}
and the expectation value $E(a,b)$ follows from%
\begin{equation}
E_{\rho}(a,b)=\sum_{A_{1},B_{2}}\int A_{1}B_{2}p(A_{1},B_{2},\lambda
)\rho(\lambda)d\lambda.  \label{18}
\end{equation}
To derive BT, one further assumption is needed, namely the factorisation of
the joint probability $p(A_{1},B_{2},\lambda)$ in terms of two independent
single particle probabilities,%
\begin{equation}
p(A_{1},B_{2},\lambda)=p(A_{1},\lambda)p(B_{2},\lambda).  \label{20}
\end{equation}
With this factorisation, condition, the expectation value takes the form%
\begin{equation}
E_{\rho}(a,b)=\int\bar{A}_{1}(\lambda)\bar{B}_{2}(\lambda)\rho(\lambda
)d\lambda,  \label{22}
\end{equation}
where%
\begin{equation}
\bar{A}_{1}(\lambda)=\sum_{A_{1}}A_{1}p(A_{1},\lambda)\text{ and\ \ }\bar {B}%
_{2}(\lambda)=\sum_{B_{2}}B_{2}p(B_{2},\lambda).  \label{24a}
\end{equation}
$\bar{A}_{1}(\lambda)$ (resp. $\bar{B}_{2}(\lambda)$) is the average over
the outcomes $A_{1}$ (resp. $B_{2}$) obtained for a fixed value of $\lambda$%
. Indeed, in its most general form, $\lambda$ does not determine the value
of a given outcome $A$, but rather the probability $p(A,\lambda)$ of
obtaining this outcome.\ This situation corresponds to \emph{stochastic }%
Bell models. The so-called ``deterministic'' Bell models appear as a particular
instance of the stochastic models when the probabilities $p(A_{1},\lambda)$
and $p(B_{2},\lambda)$ are all $0$ or $1$, in which case $\bar{A}%
_{1}(\lambda )=A_{1}(\lambda)$ and $\bar{B}_{2}(\lambda)=B_{2}(\lambda)$
meaning that a given $\lambda$ univoquely determines the value of the
measured outcomes. To obtain Bell's theorem, consider two directions $%
a,a^{\prime}$ for particle 1 measurements and two directions $b,b^{\prime}$
for particle 2 measurements (for simplicity all the directions are assumed
to be coplanar).\ Then%
\begin{equation}
\left\vert E(a,b)\mp E(a,b^{\prime})\right\vert +\left\vert
E(a^{\prime},b)\pm E(a^{\prime},b^{\prime})\right\vert \leq2V_{\max}^{2},
\label{26}
\end{equation}
where $V_{\max}$ is the maximal value that can be taken by $A$ or $B$ (here,
$\frac{1}{2}$). Eq. (\ref{26}) is easily proven \cite{bell,CS} by making use
of the factorization property (\ref{20}) within each absolute value term $%
\left\vert ...\right\vert $ and then employing triangle inequalities of the
type $\left\vert \bar{B}_{2}\mp\bar{B}_{2}^{\prime}\right\vert +\left\vert
\bar{B}_{2}\pm\bar{B}_{2}^{\prime}\right\vert \leq2V_{\max}$.

\subsection{Ruling out joint distributions}

Let's forget about hidden variables for a moment to obtain two well-known
inequalities.\ First, from the existence of a joint probability distribution
(jd) $F(A_{1},A_{1}^{^{\prime}},B_{2},B_{2}^{^{\prime}})$, it is easy to
recover the expectation values by marginalization, so that for example%
\begin{equation}
E(a,b)=\sum_{A_{1},B_{2}}A_{1}B_{2}\sum_{A_{1}^{\prime},B_{2}^{%
\prime}}F(A_{1},A_{1}^{\prime},B_{2},B_{2}^{\prime}).  \label{e27}
\end{equation}
Employing (\ref{e27}) and recalling that the absolute value of an average is
bounded by the average of the absolute values, we have%
\begin{equation}
\left\vert E(a,b)\mp E(a,b^{\prime})\right\vert \leq\sum_{A_{1}A_{1}^{\prime
}B_{2}B_{2}^{\prime}}F(A_{1},A_{1}^{\prime},B_{2},B_{2}^{\prime})\left\vert
A_{1}\left( B_{2}\mp B_{2}^{\prime}\right) \right\vert
\end{equation}
and the analog inequality for $\left\vert E(a^{\prime},b)\pm E(a^{\prime
},b^{\prime})\right\vert $.\ Adding both inequalities yields
\begin{align}
\left\vert E(a,b)\mp E(a,b^{\prime})\right\vert + & \left\vert E(a^{\prime
},b)\pm E(a^{\prime},b^{\prime})\right\vert \leq  \notag \\
&
\sum_{A_{1}A_{1}^{\prime}B_{2}B_{2}^{\prime}}F(A_{1},A_{1}^{^{%
\prime}},B_{2},B_{2}^{^{\prime}})\left( \left\vert A_{1}\left( B_{2}\mp
B_{2}^{\prime}\right) \right\vert +\left\vert A_{1}^{\prime}\left( B_{2}\pm
B_{2}^{\prime}\right) \right\vert \right) \leq2V_{\max}^{2}\text{,}
\label{i1}
\end{align}
where the right handside is obtained by using
\begin{equation}
\left\vert A_{1}\left( B_{2}\mp B_{2}^{\prime}\right) \right\vert
+\left\vert A_{1}^{\prime}\left( B_{2}\pm B_{2}^{\prime}\right) \right\vert
\leq2V_{\max}^{2}.  \label{27}
\end{equation}

The second inequality is a quantum mechanical result valid for spin-1/2
projection operators. Let $\hat{S}_{1a},\hat{S}_{1a^{\prime }}...$ denote
the operators whose eigenvalues correspond to the spin projections $%
A_{1},A_{1}^{\prime }...=\pm V_{\max }$. A direct computation establishes
that \cite{landau}%
\begin{equation}
\left( \hat{S}_{1a}\hat{S}_{2b}\mp \hat{S}_{1a}\hat{S}_{2b^{\prime }}+\hat{S}%
_{1a^{\prime }}\hat{S}_{2b}\pm \hat{S}_{1a^{\prime }}\hat{S}_{2b^{\prime
}}\right) ^{2}=4V_{\max }^{4}\pm \lbrack \hat{S}_{1a},\hat{S}_{1a^{\prime
}}][\hat{S}_{2b},\hat{S}_{2b^{\prime }}].  \label{e30}
\end{equation}%
This expression gives a bound for the norm of the operator between $(...)$.\
Since $\left\Vert \hat{S}\right\Vert =V_{\max }$ the norm of each commutator
is bounded by $2V_{\max }^{2},$ hence%
\begin{equation}
\left\Vert \hat{S}_{1a}\hat{S}_{2b}\mp \hat{S}_{1a}\hat{S}_{2b^{\prime }}+%
\hat{S}_{1a^{\prime }}\hat{S}_{2b}\pm \hat{S}_{1a^{\prime }}\hat{S}%
_{2b^{\prime }}\right\Vert \leq 2\sqrt{2}V_{\max }^{2},
\end{equation}%
and using the linearity of the operators and the fact that an expectation
value (denoted $\left\langle ...\right\rangle ,$ irrespective of the state)
is bounded by the norm yields%
\begin{equation}
\left\vert \left\langle \hat{S}_{1a}\hat{S}_{2b}\right\rangle \mp
\left\langle \hat{S}_{1a}\hat{S}_{2b^{\prime }}\right\rangle +\left\langle
\hat{S}_{1a^{\prime }}\hat{S}_{2b}\right\rangle \pm \left\langle \hat{S}%
_{1a^{\prime }}\hat{S}_{2b^{\prime }}\right\rangle \right\vert \leq 2\sqrt{2}%
V_{\max }^{2}.  \label{i2}
\end{equation}%
Obviously if the commutators in Eq. (\ref{e30}) vanished, then Eq. (\ref{i2}%
) would be bounded by $2$, just like the bound in BT. And in that case
quantum mechanics allows to compute probabilities for joint events.\ But there
is no joint distribution for non-commuting operators. Hence \cite%
{rastall81,fine,muynck,accardi}, writing IRQ for 'irrelevant to quantum mechanics',
one has%
\begin{equation}
\text{\emph{Bell's theorem is IRQ because it assumes joint distributions
where quantum mechanics denies it}.}  \label{irk1}
\end{equation}%
In the derivation of BT the factorization (\ref{20}) is what brings in the
existence of jd. Although (\ref{20}) is known as Bell's locality condition,
one can replace it with a \emph{non-local factorizable} condition and still
derive Bell's theorem \cite{annales}: non-local models also obey BT if they
are factorizable.

\subsection{Ruling out stochastic, then deterministic hidden variables}

In the quantum context involving the fragmentation of two spin-1/2 particles
formed in the singlet state%
\begin{equation}
\left\vert \psi \right\rangle =\frac{1}{\sqrt{2}}(\left\vert 1+\right\rangle
_{u}\left\vert 2-\right\rangle _{u}-\left\vert 1-\right\rangle
_{u}\left\vert 2+\right\rangle _{u} \label{sing}
\end{equation}%
where $u$ is any axis, the observed frequencies $F(A_{1},B_{2})$ are given
according to quantum mechanics by the probabilities%
\begin{equation}
P_{\psi }(A_{1},B_{2})=\left\vert \left\langle \psi \right\vert \left. 1%
\text{sign}(A_{1})\right\rangle _{a}\left\vert 2\text{sign}%
(B_{2})\right\rangle _{b}\right\vert ^{2}.  \label{28}
\end{equation}%
In terms of the LHV, Eqs. (\ref{17}) and (\ref{20}) imply%
\begin{equation}
F_{\rho }(A_{1},B_{2})=\int p(A_{1},\lambda )p(B_{2},\lambda )\rho (\lambda
)d\lambda  \label{30}
\end{equation}%
where $\rho $ is the distribution corresponding to the system having been
prepared in the singlet state $\left\vert \psi \right\rangle $. Eq. (\ref{30}%
) is inconsistent with stochastic Bell models. Indeed,
choosing $b=a$ in Eq. (\ref{28}) yields%
\begin{align}
P_{\psi }(A_{1}& =+\frac{1}{2},B_{2}\equiv A_{2}=A_{1}=+\frac{1}{2})=0,
\label{32a} \\
P_{\psi }(A_{1}& =-\frac{1}{2},B_{2}\equiv A_{2}=A_{1}=-\frac{1}{2})=0,
\label{32b}
\end{align}%
while for the other 2 possibilities
\begin{equation}
P_{\psi }(A_{1}=\pm \frac{1}{2},A_{2}=\mp \frac{1}{2})=\frac{1}{2}.
\label{e9}
\end{equation}%
Eqs. (\ref{32a})-(\ref{e9}) yield the single particle probabilities $P_{\psi
}(A_{i})=1/2$. Now summing Eq. (\ref{30}) over $A_{1}$ or $B_{2}=A_{2}$
gives
\begin{equation}
F_{\rho }(A_{i}=\pm \frac{1}{2})=\int p(A_{i}=\pm \frac{1}{2},\lambda )\rho
(\lambda )d\lambda =\frac{1}{2}\text{ \ }(i=1,2).
\end{equation}%
But since $F_{\rho }(A_{1},A_{2}=-A_{1})$ should also match (\ref{e9}), the
expressions of the type $F_{\rho }(A_{1})-F_{\rho }(A_{1},A_{2}=-A_{1})$
vanish, from which it follows that
\begin{equation}
p(A_{1}=\pm \frac{1}{2},\lambda )(1-p(A_{2}=\mp \frac{1}{2},\lambda ))=0
\end{equation}
for any $\lambda \in \Lambda $, compatible only with unit or vanishing
probability functions. Hence stochastic HV must be ruled out.

The only possibility is thus that of deterministic HV, but these must be
ruled out as well.\ The argument goes back to Wigner \cite{wigner}, and is
based on simple set theoretic assumptions -- a set $\Lambda $ over which a
probability measure is defined is partitioned into different subsets having
non-empty intersections \cite{ghirardi marinatto08}. A subset of $\Lambda $
is in correspondence with an event, so that the measure of a subset
represents the probability of the event. For example let $\Lambda _{+a}$
denote the subset such that $A_{1}(\lambda )=+\frac{1}{2}$ (and hence $%
A_{2}(\lambda )=-\frac{1}{2}$; conversely for $\lambda \in \Lambda _{-a}$ we
have $A_{2}(\lambda )=+\frac{1}{2}$ since the index is always relative to
particle 1). These subsets cover the state space $\Lambda $ such that%
\begin{equation}
\Lambda =\Lambda _{+a}\cup \Lambda _{-a}.  \label{e10}
\end{equation}%
Eq. (\ref{e10}) must be valid for any direction $a$ so that if $a^{\prime }$
denotes an arbitrary axis, we have%
\begin{equation}
\Lambda _{\pm a}=(\Lambda _{\pm a}\cap \Lambda _{+a^{\prime }})\cup (\Lambda
_{\pm a}\cap \Lambda _{-a^{\prime }}).  \label{55}
\end{equation}%
Assume now that $B_{2}$ has been measured and the outcome is known, say $%
B_{2}=-\frac{1}{2}$. The quantum mechanical probabilities,%
\begin{equation}
P_{\psi }(A_{1}=\pm \frac{1}{2},B_{2}=-\frac{1}{2})=\left\{
\begin{tabular}{l}
$\frac{1}{2}\cos ^{2}\frac{\theta _{b}-\theta _{a}}{2}$ if $A_{1}=+\frac{1}{2%
}$ \\
$\frac{1}{2}\sin ^{2}\frac{\theta _{b}-\theta _{a}}{2}$ if $A_{1}=-\frac{1}{2%
}$%
\end{tabular}%
\ \ \ \ \ \ \ \ \ \ \ \right. ,  \label{59}
\end{equation}%
should match the frequency predicted by LHV%
\begin{equation}
F_{\rho }(A_{1}=\pm \frac{1}{2},B_{2}=-\frac{1}{2})=\int_{\Lambda
_{+b}}p(A_{1},\lambda )\rho (\lambda )d\lambda ,  \label{60n}
\end{equation}%
Since $p(A_{1},\lambda )$ is $1$ or $0$ depending on whether $\lambda \in
\Lambda _{\pm a}$, Eq. (\ref{60}) becomes%
\begin{equation}
F_{\rho }(A_{1}=\pm \frac{1}{2},B_{2}=-\frac{1}{2})=\int_{\Lambda _{+b}\cap
\Lambda _{\pm a}}\rho (\lambda )d\lambda \equiv \mathcal{M}_{\Lambda
_{+b}\cap \Lambda _{\pm a}},  \label{60b}
\end{equation}%
where $\mathcal{M}_{\Lambda _{+b}\cap \Lambda _{\pm a}}$ defines the mesure
of the subset $\Lambda _{+b}\cap \Lambda _{\pm a}\subset \Lambda $ (actually
one can show that the rotational symmetry of the singlet state imposes that $%
\rho (\lambda )$ must be uniform and $\mathcal{M}$ is simply the relative
volume of $\Lambda _{+b}\cap \Lambda _{\pm a}$). If we measure\ particle 1's
spin along $a^{\prime },$ rather than along $a$, $F_{\rho }(A_{1}^{\prime
}=\pm \frac{1}{2},B_{2}=-\frac{1}{2})=\mathcal{M}_{\Lambda _{+b}\cap \Lambda
_{\pm a^{\prime }}}$. Using $\Lambda =\Lambda _{+a^{\prime }}\cup \Lambda
_{-a^{\prime }}$ [Eq. (\ref{e10})], we note that%
\begin{equation}
\Lambda _{+b}\cap \Lambda _{+a}=\left( \Lambda _{+a}\cap \Lambda _{+b}\cap
\Lambda _{+a^{\prime }}\right) \cup \left( \Lambda _{+a}\cap \Lambda
_{+b}\cap \Lambda _{-a^{\prime }}\right)
\end{equation}%
so that
\begin{equation}
\mathcal{M}_{\Lambda _{+a}\cap \Lambda _{+b}\cap \Lambda _{+a^{\prime }}}=%
\mathcal{M}_{\Lambda _{+a}\cap \Lambda _{+b}}-\mathcal{M}_{\Lambda _{+a}\cap
\Lambda _{+b}\cap \Lambda _{-a^{\prime }}}.  \label{62n}
\end{equation}%
Using the trivial inequalities $\mathcal{M}_{\Lambda _{+a}\cap \Lambda
_{+b}\cap \Lambda _{-a^{\prime }}}\leq \mathcal{M}_{\Lambda _{+a}\cap
\Lambda _{-a^{\prime }}}$ and $\mathcal{M}_{\Lambda _{+a}\cap \Lambda
_{+b}\cap \Lambda _{+a^{\prime }}}\leq \mathcal{M}_{\Lambda _{+a^{\prime
}}\cap \Lambda _{+b}}$ we infer from Eq. (\ref{62n}) that%
\begin{equation}
\mathcal{M}_{\Lambda _{+a^{\prime }}\cap \Lambda _{+b}}\geq \mathcal{M}%
_{\Lambda _{+a}\cap \Lambda _{+b}\cap \Lambda _{+a^{\prime }}}\geq \mathcal{M%
}_{\Lambda _{+a}\cap \Lambda _{+b}}-\mathcal{M}_{\Lambda _{+a}\cap \Lambda
_{-a^{\prime }}}  \label{63}
\end{equation}%
which, following (\ref{60b}) is an inequality corresponding to the
probabilities predicted by the deterministic HV. However, this inequality is
inconsistent with the quantum mechanical probabilities $P_{\psi }$: indeed
according to Eq. (\ref{59}), plugging in the $P_{\psi }$ in Eq. (\ref{63})
would lead to%
\begin{equation}
\cos ^{2}\frac{\theta _{b}-\theta _{a^{\prime }}}{2}\geq \cos ^{2}\frac{%
\theta _{b}-\theta _{a}}{2}-\sin ^{2}\frac{\theta _{a}-\theta _{a^{\prime }}%
}{2},  \label{65}
\end{equation}%
a relation that is not valid in general (eg it doesn't hold if we choose
coplanar angles obeying $0\leq \theta _{b}<\theta _{a}<\theta _{a^{\prime
}}\leq \pi /2$). Therefore, assuming deterministic HV leads to a
contradiction, so that they must be ruled out as well; therefore%
\begin{equation}
\text{\emph{Bell's theorem is IRQ because neither stochastic nor
deterministic factorizable LHV are consistent with quantum probabilities.}}
\label{irk2}
\end{equation}%
This inconsistency is grounded on the quantity $\Lambda _{+a}\cap \Lambda
_{+b}\cap \Lambda _{+a^{\prime }}$, which does not correspond to any
quantum-mechanical probability or associated quantity but is meaningful
within the Bell-type deterministic models: this is the support for the joint
events mentioned above, so that (\ref{irk2}) appears as a consequence of (%
\ref{irk1}).

\section{Locality and non-Bell-type models}

The conclusion to be drawn from the preceding Section is that factorizability, implying
the existence of joint distributions, is the origin of the inadequacy of
Bell-type LHV models to account for quantum probabilities and expecation
values.\ Non-local models can be factorizable (in which case they are also
constrained by BT) or not.\ But what about \emph{local} models? Since it is usually stated
that Eq. (\ref{20}) is the consequence of locality,
it would appear that non-factorizable models cannot be local \footnote{Bell was
actually more precise -- he carefully argued that Eq. (\ref{20}) could be derived by
assuming \emph{local causality}, that is physical theories in which the measurement outcomes can be fully
specified in terms of a complete set of beables \cite{bell-loc}. This is stronger than requiring simple locality, which only involves the absence of action at a distance and not the issues of completeness and determinism.}. In this
section we challenge this assertion by giving an overview of two different
types of models.\ The first model is built from the remark that the
purported non-locality actually arises by the combination of non-commutative
observables (precluding factorizability) and a conservation law (imposed by
rotational invariance), so giving priority to non-locality or asserting that
a conservation law is all that is needed becomes a matter of taste. The
second model, based on particle and field configurations, defuses the EPR
dilemma from the start: this model negates the existence of an element of
reality from the possibility of making a prediction with unit-probability.
As a consequence the model does not allow to efficiently complete quantum mechanics by a fully
deterministic model ascribing sub-quantum probabilities.

\subsection{Model 1: Conservation laws, holism or non-locality?}

The model described in details elsewhere (see \cite{matzkin-jpa08} and in
particular Sec.\ IV of \cite{matzkin-pra08}) is based on ensemble properties
of classical angular momenta distributions. Consider the fragmentation of an
initial particle with a total angular momentum $\mathbf{J}_{T}=0$ into 2
particles carrying angular momenta $\mathbf{J}_{1}$ and $\mathbf{J}_{2}$.
Conservation of the total angular momentum imposes $J_{1}=J_{2}\equiv J$ and%
\begin{equation}
\mathbf{J}_{1}+\mathbf{J}_{2}=0.  \label{e1}
\end{equation}%
Without further constraints (or additional knowledge), the classical
distribution in the 2-particle phase space is given by%
\begin{equation}
\rho (\Omega _{1},\Omega _{2})=N\delta (\mathbf{J}_{1}+\mathbf{J}_{2})\delta
(J_{1}^{2}-J^{2}),  \label{9}
\end{equation}%
where $N$ is a normalization constant. The corresponding distributions of
the angular momenta in physical space -- easier to visualize than $\rho $ --
is uniform on the angular momentum sphere, with $\mathbf{J}_{1}$ and $%
\mathbf{J}_{2}$ pointing in opposite directions. We can take $J=1$ without
loss of generality \cite{matzkin-jpa08}. The detectors contain a random
interaction and only deliver the results $\pm \frac{1}{2}$. Let us take a
closer look at the measurement process for a single particle whose $\mathbf{J%
}$ distribution is $\rho _{a+}$, a uniform distribution on the hemisphere
characterized by $J_{a}>0$. Let $R_{b}=\pm \frac{1}{2}$ denote the outcomes
for measurements along $b$. The system-apparatus interaction is assumed to
verify the following property: the average over the outcomes $R_{b}$ is
equal to the mean value of the projection $J_{b}$ over the initial
distribution:%
\begin{equation}
\left\langle R_{b}\right\rangle _{\rho _{a+}}=\sum_{k}kP(R_{b}=k,\rho
_{a+})=\left\langle J_{b}\right\rangle _{\rho _{a+}}=\frac{1}{2}\cos \left(
\theta _{b}-\theta _{a}\right) .  \label{e25}
\end{equation}%
Three interesting properties follow.\ (i) Eq. (\ref{e25}) along with normalization
is sufficient to impose the probabilities  $P(R_{b}=\pm
\frac{1}{2},\rho _{a+})=(\cos \left( \theta _{b}-\theta _{a}\right) \pm 1)/2$%
. (ii) Eq. (\ref{e25}) is inconsistent with the existence of elementary
probabilities depending on $\mathbf{J}$, ie there can be no $p(R_{b}=\pm
\frac{1}{2},\mathbf{J})$ such that
\begin{equation}
P(R_{b}=\pm \frac{1}{2},\rho _{a+})=\int p(R_{b}=\pm \frac{1}{2},\mathbf{J}%
)\rho _{a+}(\mathbf{J})d\mathbf{J,}
\end{equation}%
as only ensemble-dependent elementary probabilities $p(R_{b}=\pm \frac{1}{2},%
\mathbf{J},\rho _{a+})$ are consistent with Eq. (\ref{e25}) \cite%
{matzkin-jpa08,matzkin-pra08}. (iii) putting $b=a$ in Eq. (\ref{e25})
gives $P(R_{a}=\pm \frac{1}{2},\rho _{a+})=1$ or $0$; taking into account
the ensemble dependency, this means that $R_{a}=1/2\Leftrightarrow J_{a}>0\ $%
for \emph{every} $J_{a}\in \rho _{a+}$: when the distribution and
measurement axes coincide, there is no interaction and the measurement
device senses at most one hemisphere. This model is compatible eg with a
particle following a stochastic motion with its angular momentum constrained
to remain in the ensemble, the timescale of the measurement being
significantly larger than the timescale of the stochastic motion.

We now return to the 2-particle problem with the uniform distribution $\rho $%
. Eq. (\ref{e25}) becomes $\left\langle R_{ia}\right\rangle _{\rho
}=\left\langle J_{ia}\right\rangle _{\rho }=0$ where $i=1,2$ and $a$ is any
axis. Eq. (\ref{e1}) and point (iii) above imply that the outcomes and the
distributions for the particles along the same axis must be anti-correlated
along any axis $a$%
\begin{equation}
\left\langle J_{2a}\right\rangle _{\rho _{a\mp }}\equiv R_{2a}=-R_{1a}\equiv
-\left\langle J_{1a}\right\rangle _{\rho _{a\pm }}  \label{33}
\end{equation}%
where the efficient distribution $\rho _{a\pm }$ in which the particle
undergoes its stochastic motion depends on the initial position of the
angular momentum and on the choice of the measurement axis. Measuring $%
R_{1a} $ links the outcome to one of the two ensembles $\rho _{1a\pm }$
depending on whether $R_{1a}=\pm 1/2$. Note that contrarily to the
correlation between individual phase-space positions (for which one has $%
J_{2a}=-J_{1a}$ and $J_{2b}=-J_{1b}$ jointly for any axes $a$ and $b$), Eq. (%
\ref{33}) cannot hold jointly along several directions (this is a
consequence of the ensemble dependency, implying non-commutativity even for
a single particle). Since the measurement outcomes do not depend on the
individual phase-space positions, the average $E(a,b)\equiv \left\langle
R_{1a}R_{2b}\right\rangle _{\rho }$ cannot be computed from phase-space
averages, but from the probabilities of detecting a given outcome as a
function of the distribution. $E(a,b)$ is computed from the general formula%
\begin{equation}
\left\langle R_{1a}R_{2b}\right\rangle _{\rho }=\sum_{k,k^{\prime
}=-1/2}^{1/2}kk^{\prime }P_{kk^{\prime }}\text{ with }P_{kk^{\prime
}}=P(R_{1a}=k\cap R_{2b}=k^{\prime },\rho )=P(R_{1a}=k)P(R_{2b}=k^{\prime
}|R_{1a}=k).  \label{e38}
\end{equation}%
The two particle expectation takes the form%
\begin{equation}
\left\langle R_{1a}R_{2b}\right\rangle _{\rho
}=\sum_{k=-1/2}^{1/2}kP(R_{1a}=k)\left[ \sum_{k^{\prime
}=-1/2}^{1/2}k^{\prime }P(R_{2b}=k^{\prime }|R_{1a}=k)\right] .  \label{e40}
\end{equation}%
For any particle $i$ and direction $a$, we have $P(R_{ia}=\pm \frac{1}{2}%
,\rho )=\frac{1}{2}$. The conditional probability $P(R_{2b}=k^{\prime
}|R_{1a}=k)$ is the probability of obtaining $R_{2b}=k^{\prime }$ if it
known that $R_{1a}=k$. But obtaining an outcome $R_{1a}=k$ means that the
distributions for particles 1 and 2 can be restricted to $\rho _{1a[\mathrm{%
sign}(k)]}$ and $\rho _{2a[\mathrm{sign}(-k)]}$ respectively. The
conditional probability is therefore given by%
\begin{equation}
P(R_{2b}=k^{\prime }|R_{1a}=k)=P(R_{2b}=k^{\prime },\rho _{2a[\mathrm{sign}%
(-k)]}),  \label{31b}
\end{equation}%
which is a single particle probability of the type\ given in point (i) below Eq. (\ref%
{e25}). Plugging these quantities into (\ref{e40}) leads to $E(a,b)=-\frac{1%
}{4}\cos (\theta _{b}-\theta _{a})$, the quantum mechanical result for the
singlet spin state (\ref{sing}).

The present model therefore does not abide by BT. The reason is twofold.\
First comes the ensemble dependency, enforcing not only non-commutativity,
but the impossibility of ascribing elementary probabilities. Second comes
the conservation of the angular momentum: what Eq. (\ref{33}) does is to
turn the conservation of the angular momentum over the ensembles into the
conservation of the angular momentum \emph{between} these ensembles. This
means that somehow, the particles must know what ensemble was picked by the
first measurement in order to conserve the ensemble angular momentum of the
second ensemble previous to its measurement. What is really
happening is the application of the conservation law in the context of
non-commutative measurements: contrarily to the commutative case where $%
J_{1a}=-J_{2a}$ and $J_{1b}=-J_{2b}$ can hold jointly, here $R_{1a}=-R_{2a}$
and $R_{1b}=-R_{2b}$ do not. At this point it would be possible to invoke
non-locality to explain how the angular momentum can be conserved, though
one can also uphold that conservation laws and symmetry principles are just
postulated, without the need to invoke a specific mechanism. Alternatively
it can be argued that symmetries can give rise to nonlocality, a position
leading to a holistic vision of symmetries as holding beyond a space-time
framework. Note that mechanical holistic systems -- that is two systems
maintaining a mechanical link between them -- were already known to violate the Bell
inequalities (an ad-hoc model was proposed in Ref. \cite{aerts}). Here we have given a physical model that turns out to be the classical counterpart of quantum mechanical coupled angular momenta \cite{matzkin-pra08}: the violation of the Bell inequalities
is necessary in order to conserve symmetries.

\subsection{Model 2: Quantum states as equivalence classes}

The model \cite{matzkin-f} represents a single spin-1/2 by a field-particle system composed
of a small sphere, with the position of its center in the laboratory frame
being denoted by $\mathbf{x}$ and the internal spherical variables relative
to the center of the sphere by $\mathbf{r}\equiv (r,\theta ,\phi )$. A
classical scalar field $F(\mathbf{r})$ is defined on the sphere's surface,
while the point-like particle sits still at a fixed (but unknown position)
on the sphere. As in Sec.\ 2, let $B$ denote the spin projection along an
axis $b$ making an angle $\theta _{b}$ with the $z$ axis. The outcome may
depend on the position occupied by the the field on the spherical surface
and (ii) on the position of the particle. The field $F$ is defined on the
hemispherical surface centered on a given axis, the value of the field at
any point being given by the projection of that point on the axis. Let $%
\Sigma _{+a}$ denote the positive half-sphere centered on the axis $a$
making an angle $\theta _{a}$ with the $z$ axis, and $F_{\Sigma _{+a}}$
denote the field distributed on that hemisphere. $F_{\Sigma _{+a}}(\mathbf{r}%
)$ is thus defined by%
\begin{equation}
F_{\Sigma _{+a}}(\mathbf{r})=\left\{
\begin{tabular}{l}
$\mathbf{r}\cdot \mathbf{a}/\pi R^{2}$ if $\mathbf{r}\in \Sigma _{+a}$ \\
$0$ otherwise%
\end{tabular}%
\ \ \ \ \right. ,  \label{2}
\end{equation}%
$R$ being the radius of the sphere (for simplicity we will take all the axes
to be coplanar with $z$). The mean value of $\mathbf{r}\cdot \mathbf{b}/\pi
R^{2}$ taken over $\Sigma _{+a}$ is given by
\begin{equation}
\left\langle F_{\Sigma _{+b}}+F_{\Sigma _{-b}}\right\rangle _{\Sigma
_{+a}}\equiv \int_{\Sigma _{+a}}\frac{\mathbf{r}\cdot \mathbf{b}}{\pi R^{2}}d%
\mathbf{\hat{r}}=\cos \left( \theta _{b}-\theta _{a}\right) ,  \label{3}
\end{equation}%
where $d\mathbf{\hat{r}}$ denotes the spherical surface element for a sphere
of radius $R$. The only requirement we make on the particle's position is
that it must embedded within the field: the particle cannot be in a field
free region of the sphere.

When a measurement is made we assume that the apparatus along $b$ interacts
with the field $F_{\Sigma _{+a}}.$ Let $[a+b]$ and $[a-b]$ denote the
directions lying halfway between the axes $a$ (of the distribution) and $b$
or $-b$ (of the measuring direction), with respective angles $(\theta
_{b}+\theta _{a})/2$ and $(\theta _{b}+\pi +\theta _{a})/2$. We will assume
that the field-apparatus interaction results in a \emph{rotation} of the
original pre-measurement field $F_{\Sigma _{+a}}$ toward both of the
apparatus axes, $F_{\Sigma _{+a}}\rightarrow F_{\Sigma _{+b}}+F_{\Sigma
_{-b}}$. A definite outcome $B=\pm \frac{1}{2}$ depends on which of the
hemispheres $\Sigma _{\pm b}$ the particle is after the interaction. In
terms of the field, this probability is given by the relative value of the
average of the rotated field $F_{\Sigma _{+b}}+F_{\Sigma _{-b}}$ over the
intermediate 'half-rotated' hemisphere $F_{\Sigma _{\lbrack a\pm b]}}$
depending on the initial field $F_{\Sigma _{+a}}$, yielding in accordance
with Eq. (\ref{3})%
\begin{align}
P_{\Sigma _{+a}}(B& =+\frac{1}{2})=\left\vert \left\langle F_{\Sigma
_{+b}}+F_{\Sigma _{-b}}\right\rangle _{\Sigma _{\lbrack a+b]}}\right\vert
^{2}/N=\cos ^{2}\frac{\theta _{b}-\theta _{a}}{2}  \label{10} \\
P_{\Sigma _{+a}}(B& =-\frac{1}{2})=\left\vert \left\langle F_{\Sigma
_{+b}}+F_{\Sigma _{-b}}\right\rangle _{\Sigma _{\lbrack a-b]}}\right\vert
^{2}/N=\sin ^{2}\frac{\theta _{a}-\theta _{b}}{2}  \label{12}
\end{align}%
with $N\ $being the sum of both terms. If $b$ and $a$ are taken to be the
same, then one has $\Sigma _{\lbrack a+a]}\equiv \Sigma _{+a}$ and $%
P_{\Sigma _{+a}}(A=\pm \frac{1}{2})=1$ and $0$ respectively. Hence a
field $F_{\Sigma _{+a}}$ corresponds to a well-defined positive spin projection
along the $a$ axis. In this case the symmetry axis of the field distribution
coincides with the post-measurement axis and the field-apparatus interaction
may change the position of the particle though it remains within the
hemisphere $\Sigma _{+a}$. On the other hand when $b$ and $a$ lie along
different directions, the spin projection along $b$ only acquires a value $%
B=\pm \frac{1}{2}$ \emph{after} the field has interacted with the
measurement apparatus and rotated toward the measurement axis: the
measurements do not commute, and thus joint spin measurements along
different axes are undefined.

Since fields obey the principle of superposition, we can envisage
superpositions of fields defined on different hemispheres. But fields
defined on different hemispheres turn out to be \emph{equivalent }to a field
defined on a single hemisphere. Indeed it is easy to see that one can write
for any axis $u$
\begin{equation}
F_{\Sigma _{+a}}\sim \cos (\frac{\theta _{u}-\theta _{a}}{2})F_{\Sigma
_{+u}}+\sin (\frac{\theta _{u}-\theta _{a}}{2})F_{\Sigma _{-u}},  \label{z1}
\end{equation}%
meaning that although the two fields on the right and left handsides of Eq. (%
\ref{z1}) are different -- they are not defined on the same hemispherical
surfaces --, they lead to exactly the same predictions. Indeed, when
measurements are made along \emph{any} axis $b$ the averages of the left and
right handsides (hs) of Eq. (\ref{z1}) give the same result $\cos (\frac{%
\theta _{a}-\theta _{b}}{2})$. These fields thus define an \emph{equivalence
class.} From the particle standpoint, the\ field on the rhs of Eq. (\ref{z1}%
), $F_{rhs}$ implies a different behavior: the no-perturbation axis is $u$,
not $a$, and the particle distribution is not uniform. Hence there is a
probability function $p_{F_{rhs}}(U=\pm \frac{1}{2},\mathbf{r})=1$ or $0$
depending on whether $\mathbf{r\in }\Sigma _{\pm u}$ and such that%
\begin{equation}
P_{F_{rhs}}(U=\pm \frac{1}{2})=\int p_{F_{rhs}}(U=\pm \frac{1}{2},\mathbf{r}%
)\rho _{rhs}(\mathbf{r})d\mathbf{r}=\cos ^{2}\left( \frac{\theta _{u}-\theta
_{a}}{2}+\frac{\pi }{4}(1\pm 1)\right) ,  \label{z2}
\end{equation}%
where $\rho _{rhs}(\mathbf{r})$ denotes the particle distribution when the
field is given by the rhs of Eq. (\ref{z1}). However for $b\neq u$ there is
no probability function $p_{F_{rhs}}(B=\pm \frac{1}{2},\mathbf{r})$ hence $%
P_{F_{rhs}}(B=\pm \frac{1}{2})$ cannot depend on $\mathbf{r}$: the particle
position does not ascribe probabilities and there is no sub-field mechanism
that determines the outcome.\ This is consistent with Eqs. (\ref{10})-(\ref%
{12}) in which the field rotation does not allow to define joint
probabilities of the type $P_{F_{rhs}}(U=\pm 1\cap B=\pm 1)$. Note that
measuring $A$ in the field $F_{rhs}$ involves a perturbation in which the
fields interfere due to the rotations in such a way as to obtain $%
P_{F_{rhs}}(A=-\frac{1}{2})=0$ irrespective of the initial the particle's
position. The model cannot give a more specific interpretation in terms of
the particle for this result produced by the interaction between the system
and the apparatus measuring the spin projection along $a$.

Assume now an initial two-particle system is fragmented into two subsystems
flying apart in opposite directions. Each of the two particles is embedded
in a field defined on the surface of a small sphere. $\mathbf{x}_{1}$ (resp.
$\mathbf{x}_{2}$) denotes the position of the subsystem 1 (resp.\ 2) sphere
in the laboratory frame.\ The internal variables within each sphere are
labeled by $\mathbf{r}_{1}$ and $\mathbf{r}_{2}$. As soon as the
fragmentation process is completed, the positions of each point-like
particle as well as the fields are fixed, the spin of each system
depending on the field distribution and the particle position on its
spherical surface. The correlation of the particle positions must be set as $%
\mathbf{r}_{1}=-\mathbf{r}_{2}$ in order to achieve $A_{2}=-A_{2}$ for any
axis $a$ when there is no measurement perturbation. However the main element
characterizing the correlations is the field distribution. For example the
total field arising by correlating $F_{\Sigma _{+a}}^{1}$ defined on
subsystem 1's sphere with $\Sigma _{-a}^{2}$ on subsystem 2 is given by $%
F_{\Sigma _{+a}}^{1}(\mathbf{r}_{1})F_{\Sigma _{-a}}^{2}(\mathbf{r}_{2})$.
If in addition we also require the correlation $F_{\Sigma
_{-a}}^{1}\leftrightarrow F_{\Sigma _{+a}}^{2}$ the total field is given by
the expression
\begin{equation}
F_{\aleph }(\mathbf{r}_{1},\mathbf{r}_{2})=F_{\Sigma _{+a}}^{1}(\mathbf{r}%
_{1})F_{\Sigma _{-a}}^{2}(\mathbf{r}_{2})-F_{\Sigma _{-a}}^{1}(\mathbf{r}%
_{1})F_{\Sigma _{+a}}^{2}(\mathbf{r}_{2}).  \label{58}
\end{equation}%
The definition of $F_{\aleph }$ is mathematically non-separable over the
individual subsystem spheres; this means that the field is defined as a
whole, jointly over the two spheres.\ This is the only way to account for
correlations between more than two hemispheres \footnote{Actually it is possible to
replace the non-separable field by a separable, factorizable one if the field is allowed
to take complex values \cite{matzkin-f}; the separable field is then expanded as the sum of two non-separable
fields, one of which never contributes to the averages, and the other being $F_{\aleph }$.}; without further
specifications, non-separability has nothing to do with non-locality (the
field is set at the source in the intersection of the past light-cones of
both system's space-time location). Recall that non-separable functions are
not exceptional in classical physics, eg the classical action for
multiparticle systems is non-separable, but that does not make particle
classical mechanics non-local.

Let us now investigate measurements along arbitrary directions $c$ for
particle 1 and $b$ for particle 2, and consider%
\begin{equation}
P_{\aleph }(C_{1}=1,B_{2}=1)=\frac{1}{2}|\left\langle F_{\Sigma _{\lbrack
a+c]}}^{1}\right\rangle _{+c}\left\langle F_{\Sigma _{\lbrack
-a+b]}}^{2}\right\rangle _{+b}-\left\langle F_{\Sigma _{\lbrack
-a+c]}}^{1}\right\rangle _{+c}\left\langle F_{\Sigma _{\lbrack
a+b]}}^{2}\right\rangle _{+b}|^{2}  \label{60}
\end{equation}%
where $N=2$ is the probabilities normalization factor. $P_{\aleph }$ is a
two-outcome probability and consequently depends on the correlated local
averages of the both subsystems' fields rotated by the local interaction of
each field with the measurement apparatus along the axes $c$ and $b$. It can
be simplified by using the expressions employed for the single particle
averages: we then see that the expression between $\left\vert ...\right\vert
$ reduces to $\cos (\frac{\theta _{b}-\theta _{c}}{2})$ and is independent
of $a$. Therefore $P_{\aleph }$ \emph{does not depend on the direction }$a$
of the single-particle fields that define $F_{\aleph }$ in Eq. (\ref{58}).
This implies the \emph{equivalence} between fields $F_{\Sigma
_{+a}}^{1}F_{\Sigma _{-a}}^{2}-F_{\Sigma _{-a}}^{1}F_{\Sigma _{+a}}^{2}$
defined by different directions $a$, i.e. for any $b\neq a$%
\begin{equation}
F_{\Sigma _{+a}}^{1}F_{\Sigma _{-a}}^{2}-F_{\Sigma _{-a}}^{1}F_{\Sigma
_{+a}}^{2}\sim F_{\Sigma _{+b}}^{1}F_{\Sigma _{-b}}^{2}-F_{\Sigma
_{-b}}^{1}F_{\Sigma _{+b}}^{2}.  \label{62}
\end{equation}%
Both of these fields lead exactly to the same predictions for measurements
along arbitrary axes and can thus not be distinguished. We will denote the
left and right handsides of Eq. (\ref{62}) by $F_{\aleph (a)}$ and $%
F_{\aleph (b)}$ respectively. One consequence is that when computing $%
P_{\aleph }(A_{1},B_{2})$ one can use any of the two forms (\ref{62}).
Employing $F_{\aleph (a)}$ allows to make a conditional inference for $B_{2}$%
, given that the measurement yielding $A_{1}$ does not perturb subsystem 1
and thus reveals to which hemisphere $\Sigma _{\pm a}$ $\mathbf{r}_{1}$
belonged previous to the measurement:%
\begin{align}
P_{\aleph (a)}(A_{1}=\frac{1}{2},B_{2}=\frac{1}{2})& =P(A_{1}=\frac{1}{2}%
)P(B_{2}=\frac{1}{2}|A_{1}=\frac{1}{2})=P(\mathbf{r}_{1}\in \Sigma
_{+a}^{1})P(B_{2}=\frac{1}{2}|\mathbf{r}_{1}\in \Sigma _{+a}^{1})
\label{q10} \\
& =P(\mathbf{r}_{1}\in \Sigma _{+a}^{1})P(B_{2}=\frac{1}{2}|\mathbf{r}%
_{2}\in \Sigma _{-a}^{2})=P(\mathbf{r}_{1}\in \Sigma _{+a}^{1})P_{F_{\Sigma
_{-a}}^{2}}(B_{2}=\frac{1}{2}).  \label{q11}
\end{align}%
The last step yields $P_{F_{\Sigma _{-a}}^{2}}(B_{2}=\frac{1}{2})$ which is
a single subsystem probability; this step is justified by the fact that
given $\mathbf{r}_{2}\in \Sigma _{-a}^{2}$, the field over subsystem 2 is
equivalent, as discussed below Eq. (\ref{z2}) to $F_{\Sigma _{-a}}^{2}$.
Eqs. (\ref{q10})-(\ref{q11}) can be repeated by employing $F_{\aleph (b)}$,
giving $P_{\aleph (b)}(A_{1},B_{2})$ in terms of a conditional probability
inferred from a no-perturbation measurement along $b$ (for subsystem 2). As
in the single particle system case each particular realization of an
equivalence class gives rise to different, incompatible, accounts grounded
on the measurement that does not disturb the original field. Here however
the equivalence class holds relative to the two-particle system, but
relative to a single subsystem the specific form taken for $F_{\aleph }$ has
different implications regarding the relation between the particle position
and a given outcome. This does not affect a single subsystem probabilities, $%
P(A_{i})=\frac{1}{2}$ for any $a$, but conditional probabilities can only be
explicited when the form of the field corresponds to a measurement axis.\
For example $P_{\aleph (b)}(A_{1}=\frac{1}{2}|C_{2}=\frac{1}{2})$ cannot be
computed: it is not correlated with the particle positions and due to the
subsystems-apparata interactions, no inferences can be made; but $P_{\aleph
(c)}(A_{1}=\frac{1}{2}|C_{2}=\frac{1}{2})$ can be inferred in terms of a
single subsystem probability. The situation was the same for the single
particle system described above, where no elementary probability could be
ascribed to compute $P_{F_{rhs}}(B)$.

In this model, the particles' positions thus appear as pre-determined but
only determine the outcome when there is no field perturbations from the
apparatus interaction. The field configurations can also be taken as hidden
variables and they do ascribe probabilities but only as members of an
equivalence class that does not give a more complete specification than
afforded by the quantum-mechanical state. The first implication is that
there is no pre-existing outcome as an element of reality, even when it is
possible to make a prediction with unit probability (in this case also there
is an infinity of field/particle configurations giving that outcome). The
second is that a given field/particle configuration (even if known) does not
allow to specify sub-quantum probabilities for measurements along arbitrary
axes. Any prediction that would complete quantum-mechanics, like the
inference made on one subsystem's outcome once the other outcome is known,
relies on a specific (but fictitious) field/particle configuration for which
one of those measurements does not give rise to perturbations. This model
therefore expels the locality issue from the EPR paradox (completeness of QM
\emph{or }simultaneous existence, based on locality, of physical quantities
associated with non-commuting observables): by equating a quantum state with
an equivalence class comprising an infinity of possible field-particle
configurations the model denies the simultaneous existence of those physical
quantities regardless of the locality issue, while keeping the physical
predictions invariant. Note that the idea of a quantum state as being a label
for an ensemble of underlying phenomena appears naturally when classical fields are considered
as constituting a sub-quantum level of description \cite{khrennikov05,khrennikov08}.

\section{Conclusion}

Bell-type hidden variable models give an explicit mathematical formulation
of the EPR's simultaneous reality requirement concerning incompatible
quantities.\ These models are irrelevant to quantum mechanics in so far as
the latter denies the existence of joint distributions for incompatible
quantities.\ We have argued that it is possible to uphold non-commutativity
and locality simultaneously and constructed to that effect two different
types of models. However these models despite being local fail both
Einstein's goal of efficiently \emph{completing} quantum mechanics and
Bell's goal of implementing explicitly \emph{causality} to describe quantum
correlations.

\end{document}